\documentclass[
prd,
superscriptaddress,
tightenlines,
showkeys,
nofootinbib,
aps,
amsfonts,
amssymb,
twocolumn
]{revtex4}

\usepackage{color}
\usepackage{multirow} 
\usepackage{amsmath}
\usepackage{graphicx}
\usepackage{mathrsfs}
\usepackage{epic}
\usepackage{eepic}
\usepackage{epsfig}
\usepackage{latexsym}
\usepackage{float}
\usepackage[all]{xy}
\usepackage{amsfonts}
\usepackage{amssymb} 
\usepackage{xcolor}
\usepackage{multirow,array}
\usepackage{tikz} 
\usepackage{shorthand,slashed}
\usepackage[
    colorlinks,
    linkcolor=blue!75!white,           
    citecolor=red,            
    filecolor=blue!75!white,        
    urlcolor=blue!75!white,          
    hyperfootnotes]{hyperref}

\usepackage[paperwidth=210mm,paperheight=297mm,centering,hmargin=1.9cm,vmargin=2.5cm]{geometry}

\renewcommand{\lq}{\ell_q}
\newcommand{\ubr}[1]{\raisebox{1.5ex}{\hspace{#1ex}$\frown$\relax}}
\newcommand{\lbr}[1]{\raisebox{-1.5ex}{\hspace{#1ex}$\smile$\relax}}

\usepackage[caption=false]{subfig}

\begin{document}
\preprint{HRI-RECAPP-2015-013, MITP/15-043}

\title{\hfill{}{\footnotesize\sf HRI-RECAPP-2015-013, MITP/15-043}\\Pair Production of Scalar Leptoquarks at the LHC to NLO Parton Shower Accuracy}

\author{Tanumoy Mandal}
\email{tanumoy.mandal@physics.uu.se}
\affiliation{Regional Centre for Accelerator-based Particle Physics,
Harish-Chandra Research Institute, Chhatnag Road, Jhusi, Allahabad  211 019, India}
\affiliation{High Energy Physics Division,
Department of Physics and Astronomy,
Uppsala University,
Box 516, SE-75120 Uppsala, Sweden}

\author{Subhadip Mitra}
\email{subhadip.mitra@iiit.ac.in}
\affiliation{Center for Computational Natural Sciences and Bioinformatics,
International Institute of Information Technology, Hyderabad 500 032, India}

\author{Satyajit Seth}
\email{sseth@uni-mainz.de}
\affiliation{PRISMA Cluster of Excellence, Institut f\"{u}r Physik,
Johannes Gutenberg-Universit\"{a}t Mainz, 
D\,-\,55099 Mainz, Germany }

\begin{abstract} 

We present the scalar leptoquark pair production process at the LHC computed at the next-to-leading order in QCD, matched to the 
{\sc Pythia} parton shower using the MC@NLO formalism. We consider the leading order decay of a leptoquark to a lepton ($e,\m,\ta$ or $\n_e,\n_\m,\n_\ta$) and a jet and 
observe the effects of parton shower on the final states. 
For demonstration, we display the kinametical distributions of a selection of observables along with their scale
uncertainties for the 13 TeV LHC. We also present pair production 
cross sections and $K$-factors using massless five-quark flavor scheme for different LHC center-of-mass energies.  The complete
stand-alone code is available online.

\end{abstract}

\pacs{14.80.Sv, 23.20.Ra, 12.38.-t}
\keywords{Leptoquark, Pair production, NLO computation, LHC}

\maketitle

\section{Introduction}

Leptoquark (LQ or $\lq$) is the generic name of bosons (scalars or vectors) that can couple
to a quark and a lepton simultaneously. They appear in different beyond the Standard 
Model (BSM) scenarios like the models with quark lepton compositeness \cite{Schrempp:1984nj}, Pati-Salam models \cite{Pati:1974yy}, $SU(5)$ grand unified theories \cite{Georgi:1974sy}, the colored Zee-Babu model \cite{Kohda:2012sr}, $R$-parity violating supersymmetric models \cite{Barbier:2004ez} etc. 
The LHC is looking for their signatures \cite{Aad:2011ch,ATLAS:2012aq,ATLAS:2013oea,Aad:2015caa,Chatrchyan:2012vza,Chatrchyan:2012st,CMS:zva,Khachatryan:2014ura,CMS:2014qpa,Khachatryan:2015vaa,CMS:2015kda}. 

Recently, CMS has searched for the first two generations of scalar LQs at the 8 TeV LHC with 19.7 fb$^{-1}$ of integrated luminosity in two different channels -- $\hat\ell\hat\ell jj$ and $\hat\ell\n_{\hat\ell} jj$~\cite{Khachatryan:2015vaa} (also see Refs. \cite{CMS:2014qpa} and \cite{CMS:zva}) (we use $\hat\ell$ to denote a charged lepton from the first two generations and $\ell$ for any charged lepton, i.e., $\hat\ell=e^\pm,\m^\pm$ and $\ell=e^\pm,\m^\pm,\ta^\pm$). Considering only LQ pair production, the analysis puts 95\% C.L. mass exclusion limits (ELs) on the first generation LQ at $M_{\lq} = 1005\ (845)$ GeV for $\bt = 1\ (0.5)$, where $\bt$ is the branching fraction for a LQ to decay to an electron-quark pair and $M_{\lq}$ denotes the mass of LQs.  For the second generation, the corresponding limits are put at 1080 (760) GeV. CMS also excludes third generation LQs decaying to a $\ta$ and a $b$-quark with $\bt=1$ up to 740 GeV~\cite{CMS:2014qpa}. In the first generation search, mild excesses of events compared to the Standard Model (SM) background in both channels for LQs with mass around 650 GeV  were observed. Currently, these excesses have attracted considerable attention in the  literature \cite{Varzielas:2015iva,Queiroz:2014pra,Allanach:2015ria,Dhuria:2015hta,Allanach:2014nna,Chun:2014jha,Bai:2014xba,
Dutta:2015dka,Evans:2015ita,Dey:2015eaa,Berger:2015qra}. 

For the first generation LQ, a more stringent limit comes from ATLAS. With 20 fb$^{-1}$ of integrated luminosity at the 8 TeV LHC, ATLAS rules out first (second) generation LQs decaying to an $e$ (a $\m$) and a jet with $\bt=1$ up to 1050 (1000) GeV~\cite{Aad:2015caa}. For the third generation, ATLAS rules out LQs decaying to a $b$-quark and a $\n_\ta$ with $\bt=0$ up to 625 GeV whereas for LQs decaying to a $t$-quark and a $\n_\ta$ with $\bt=0$ the excluded range is between 210 and 640 GeV~\cite{Aad:2015caa}.

Most of the experimental analyses use the fixed order (FO) result for the LQ pair production cross section computed at the next-to-leading order (NLO) of QCD~\cite{Kramer:2004df}.
At FO, {${\cal{O}}(\alpha_{s})$} contributions consist of real QCD emissions from the Born subprocesses 
and the interference between the Born diagrams and the one-loop corrected Born diagrams. 
However, to get a more realistic picture of the 
final state particles, one has to combine the FO NLO results with parton showers (PS) that consistently resum large logarithmic terms in the collinear 
limit, thereby expanding the coverage of the kinematical region. 
In this paper, we compute the pair production of scalar LQs at the NLO of QCD accuracy; i.e., we analyze the {${\cal{O}}(\alpha_{s})$} corrections 
to the following leading order (LO) partonic subprocess at the LHC, 
\ba 
\left.\begin{array}{c}
q\bar q\\
~\\
gg
\end{array} 
\begin{array}{c}
\searrow \\\nearrow
\end{array}
\left(\lq\,{\lq}\right)
\begin{array}{c}
\nearrow\\\to \\\searrow
\end{array}
\begin{array}{c}
\ell j \ubr{-2.0}\, \ell j\lbr{-2.0}\\
\ell j \ubr{-2.0}\, \n j\lbr{-2.0}\\
\n   j \ubr{-2.0}\, \n j\lbr{-2.0}
\end{array}\right\}
,
\ea 
where the curved connections above or below mark a pair of a lepton ($\ell$ or $\n=\n_e,\n_\m,\n_\ta$) and a jet ($j$) coming from the decay of a LQ, retaining all the spin-correlation effects at the LO accuracy and then we match the fixed order result with the {\sc Pythia} PS.  
Computation of such multijet inclusive event samples is important to properly interpret experimental data.\footnote{Not only the PS but inclusion of other processes with similar final states can also affect the ELs significantly --- see Ref.~\cite{Mandal:2015vfa} to see how inclusive single production events in the pair production signal can affect the ELs significantly. Combining single and pair productions in signal simulations leads to a more realistic signal~\cite{Belyaev:2005ew,Mandal:2012rx} estimation in general.} Compared to FO calculations, matching with the PS improves the accuracy of theoretical estimations of various kinematic distributions. These improved and more precise distributions can play a crucial role if advanced techniques like multivariate analysis are employed to separate LQ signal from SM background or differentiate LQs from other BSM candidate with similar signatures (like leptogluons for example). 

While matching the FO NLO results with PS, we use the MC@NLO formalism \cite{Frixione:2002ik} 
that avoids double counting in the hard and collinear regions at the NLO level by adding suitable counterterms.
As our NLO+PS computation is performed using the {\sc MadGraph}5\_{\sc aMC@NLO} framework \cite{Alwall:2014hca},
it is flexible to any choice of cuts, scales, parton distribution functions (PDFs),  etc. 
We use this paper to elaborate the details of the computation and demonstrate its reliability and applicability.
Hence, it will facilitate the experimentalists to generate more realistic QCD NLO+PS level events in future.

\begin{figure}
\includegraphics[width=0.5\linewidth]{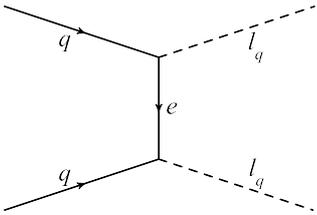}
\caption{The LO lepton exchange diagram for pair production of first generation LQ.}\label{fig:lep_ex}
\end{figure}

\section{NLO+PS Computation}

We use the same parametrization as in Ref.~\cite{Mandal:2015vfa} and consider the following simplified models for scalar LQs with electromagnetic charges, $-1/3$ and $2/3$ respectively,
\begin{eqnarray}
\mathscr L_{\rm int}^{\rm A} &=& \lm_{\ell} \left(\sqrt{\eta_{\rm L}}\, \bar u_{\rm R} \ell^+_{\rm L} + \sqrt{\eta_{\rm R}}\, \bar u_{\rm L} \ell^+_{\rm R}\right)\lq+ \lm_{\nu } \bar d_{\rm R} \tilde \nu_{\rm L}^\ell \lq\nn\\ 
&& +\ {\rm H.c.},\label{eq:modelA}\\
\mathscr L_{\rm int}^{\rm B} &=& \lm_{\ell} \left(\sqrt{\eta_{\rm L}}\, \bar d_{\rm R} \ell^-_{\rm L} + \sqrt{\eta_{\rm R}}\, \bar d_{\rm L} \ell^-_{\rm R}\right)\lq + \lm_{\nu } \bar u_{\rm R} \nu_{\rm L}^\ell \lq\nn\\
&& +\ {\rm H.c.},\label{eq:modelB}
\end{eqnarray}
where, 
\ba
\lm_{\ell}^2 = \bt\lm^2,\quad  \lm_{\n }^2 = \lt(1-\bt\rt)\lm^2\;, 
\ea
assuming there is no unknown decay mode of the LQs and there is no mixing among different generations. Here  $\eta_{\rm L}$ and $\eta_{\rm R}=(1-\eta_{\rm L})$ are the charged lepton chirality fractions, i.e., $\eta_{\rm L} \left(\eta_{\rm R}\right)$ gives the fraction of charged leptons coming from a LQ decay that are left handed (right handed). We set $\eta_L$ = 1 in our computation.\footnote{In general, the LHC is insensitive to this parameter. However, in case of a third generation LQ that couples to a $t$-quark, it might be accessible via the spin of the $t$-quark.} By parameter rescaling, the above models could be connected easily to the ones generally found in the literature (see e.g., Refs. \cite{Blumlein:1992ej,Hewett:1997ce}).

For all three generations of LQs, we implement Model A in the FeynRules package \cite{Alloul:2013bka} together with the SM
in order to collect all the tree level couplings and fields in  Universal FeynRules Output (UFO) format. Then, we renormalize the Lagrangians to get rid of the UV divergences 
that appear in loop calculation. It is well known that any one-loop amplitude can be written as linear combinations of scalar tadpole, bubble, triangle and box integrals together with rational terms that 
come from the ($d-4$) part of a $d$-dimensional integral. Within the framework of the Ossola, Papadopoulos, and Pittau (OPP) reduction technique \cite{Ossola:2006us}, the ($d-4$) part that originates form the denominator of the integrand is called the 
$R_1$ term, whereas the other rational term $R_2$ comes due to the ($d-4$) component of the numerator of such an integrand. 
We make use of the {\sc NLOCT} package \cite{Degrande:2014vpa} to calculate the UV counterterms within the on-shell renormalization scheme and to determine the rational 
$R_2$ terms required for the OPP reduction. The $R_1$ terms can be calculated as a four-dimensional integration using a particular set of scalar integrals \cite{Ossola:2008xq}. 
Like Ref. \cite{Kramer:2004df}, we also neglect the lepton exchange diagram (see Fig. \ref{fig:lep_ex}) that depends on the LQ-lepton-quark coupling ($\lm$) and contributes at $\mc O(\lm^4)$ in the cross section. For the first generation LQs, we estimated its contribution to be $\lesssim$ 10\% of the LO QCD mediated contribution for $\lm$ as large as 0.5 at the 8 TeV LHC~\cite{Mandal:2015vfa}. For higher generations, it would be even smaller because of the relative suppression of the initial PDFs.
Hence, all our cross section estimations are essentially model independent and are equally applicable for Model B too. An analysis for the $\ell \ell jj$ channel might also be used for a LQ with charge $5/3~(-4/3)$ that couples to a $u$-type ($d$-type) quark and a charged lepton (but not to a neutrino)~\cite{Mandal:2015vfa} as long as $\ell^\pm$'s are not distinguished. However, only for the third generation, we implement Model B separately since a $t$-quark decays inside the detector and leaves very different signatures from the $b$-quark.

As already mentioned, the complete computational setup for the NLO+PS correction of the LQ pair production is prepared under the automated {\sc MadGraph}5\_{\sc aMC@NLO} environment~\cite{Alwall:2014hca}.
We use {\sc MadGraph}5 \cite{Alwall:2011uj} for the Born level computations. The real emission corrections are 
computed with {\sc MadFKS} \cite{Frederix:2009yq} which uses the  subtraction scheme of Frixione, Kunszt, and Signer (FKS) \cite{Frixione:1995ms}.
To evaluate the virtual contribution, we use {\sc MadLoop} \cite{Hirschi:2011pa} which relies on the OPP reduction scheme. 
After the generation of NLO events, the decay of each LQ into a lepton-jet pair is organized by {\sc MadSpin} 
\cite{Artoisenet:2012st}, which takes care of decays of heavy resonances preserving all spin information at the tree level accuracy. 
The NLO events, thus decayed, are then matched to the {\sc Pythia8} \cite{Sjostrand:2014zea} PS following the MC@NLO formalism \cite{Frixione:2002ik}. The interfacing between different packages is made automated in {\sc aMC@NLO} \cite{Torrielli:2010aw}. 

We fix the electroweak input parameters ({\em i}) $M_{\rm Z} = 91.188$ GeV, 
({\em ii}) $G_{\rm F} = 16.637\times 10^{-6}$ GeV$^{-2}$ and ({\em iii}) $\alpha_{\rm EM}^{-1} = 132.237$ in our computations. 
The electroweak mixing angle and the mass of the $W^{\pm}$-boson are calculated from these three independent input parameters.
For the PDFs we use MSTW(n)lo2008cl68 sets \cite{Martin:2009iq} that determine the value of strong coupling $\alpha_{s}(M_Z)$ at (N)LO in QCD and use $n_f = 5$ 
massless quark flavors, i.e., the bottom quark is also treated effectively as a massless quark, since its running mass decreases significantly as the hard scale increases. 
Unless  otherwise stated, we set $M_{\lq} = 650$ GeV with $\lambda = 0.3$.

The central values of the renormalization ($\mu_R$) and factorization ($\mu_F$) scales are set equal to $M_{\lq}$, 
the standard choice \cite{Kramer:2004df, CMS:2014qpa}. 
To estimate the scale uncertainties of various 
observables, we set 
\ba 
\mu_R = \mu_F = \zeta\,M_{\lq}\,,\label{eq:scales}
\ea 
with $\zeta = \{1/2, 1,2\}$. 

Events are generated with extremely loose cuts to get unbiased results. At the time of showering these events with {\sc Pythia8}, jets are 
clustered using the anti-$k_{\rm T}$ algorithm \cite{Cacciari:2008gp} with radius parameter $r=0.5$. 
Before we present the NLO+PS results in the next section, we note two points in favor of the validity of our setup:
\emph{(i)} we find exact numerical cancellation between the double and single poles coming from the 
real and virtual parts and \emph{(ii)} our FO NLO results are in extremely good agreement with the NLO results already 
available \cite{Kramer:2004df}.

\begin{table}[]
\bc
\begin{tabular}{|c|lr|}
\hline
$M_{\lq}$(TeV)&$\sg_{NLO}$ (fb)&$K$-factor\\\hline\hline
 &\multicolumn{2}{c|}{~~~~~~~~MSTW(n)lo2008cl68 PDFs~~~~~~~~}\\\cline{2-3}
0.2 	&$( 7.757\pm 0.082)\times 10^{4}$	& 1.363 \\
0.4 	&$( 2.295\pm 0.019)\times 10^{3}$	& 1.375 \\
0.6 	&$( 2.292\pm 0.018)\times 10^{2}$	& 1.372 \\
0.8 	&$( 3.829\pm 0.033)\times 10^{1}$	& 1.376 \\
1.0 	&$( 8.190\pm 0.065)\times 10^{0}$	& 1.336 \\
1.2 	&$( 2.113\pm 0.015)\times 10^{0}$	& 1.315 \\
1.4 	&$( 6.017\pm 0.046)\times 10^{-1}$	& 1.280 \\
1.6 	&$( 1.907\pm 0.018)\times 10^{-1}$	& 1.295 \\
1.8 	&$( 6.253\pm 0.052)\times 10^{-2}$	& 1.292 \\
2.0 	&$( 2.116\pm 0.014)\times 10^{-2}$	& 1.275 \\\cline{2-3}
-----&\multicolumn{2}{c|}{CTEQ6(M/L1) PDFs }\\\cline{2-3}
0.2 	&$( 7.518\pm0.083)\times 10^{4}$	& 1.491\\
0.4 	&$( 2.217\pm0.018)\times 10^{3}$	& 1.577\\
0.6 	&$( 2.233\pm0.018)\times 10^{2}$	& 1.649\\
0.8 	&$( 3.781\pm0.032)\times 10^{1}$	& 1.724\\
1.0 	&$( 8.245\pm0.071)\times 10^{0}$	& 1.742\\
1.2 	&$( 2.210\pm0.015)\times 10^{0}$	& 1.810\\
1.4 	&$( 6.548\pm0.059)\times 10^{-1}$	& 1.852\\
1.6 	&$( 2.108\pm0.020)\times 10^{-1}$	& 1.919\\
1.8 	&$( 7.113\pm0.053)\times 10^{-2}$	& 1.988\\
2.0 	&$( 2.486\pm0.018)\times 10^{-2}$	& 2.038\\\hline
\end{tabular}
\caption{NLO cross sections and $K$-factors computed in two different PDFs for the 14 TeV LHC.}\label{tab:compare}
\ec
\end{table}

\section{Numerical Results}

\begin{figure*}
\bc
\begin{tabular}{m{0.4\linewidth}m{0.1\linewidth}m{0.4\linewidth}}
\subfloat[]{\includegraphics[width=\linewidth]{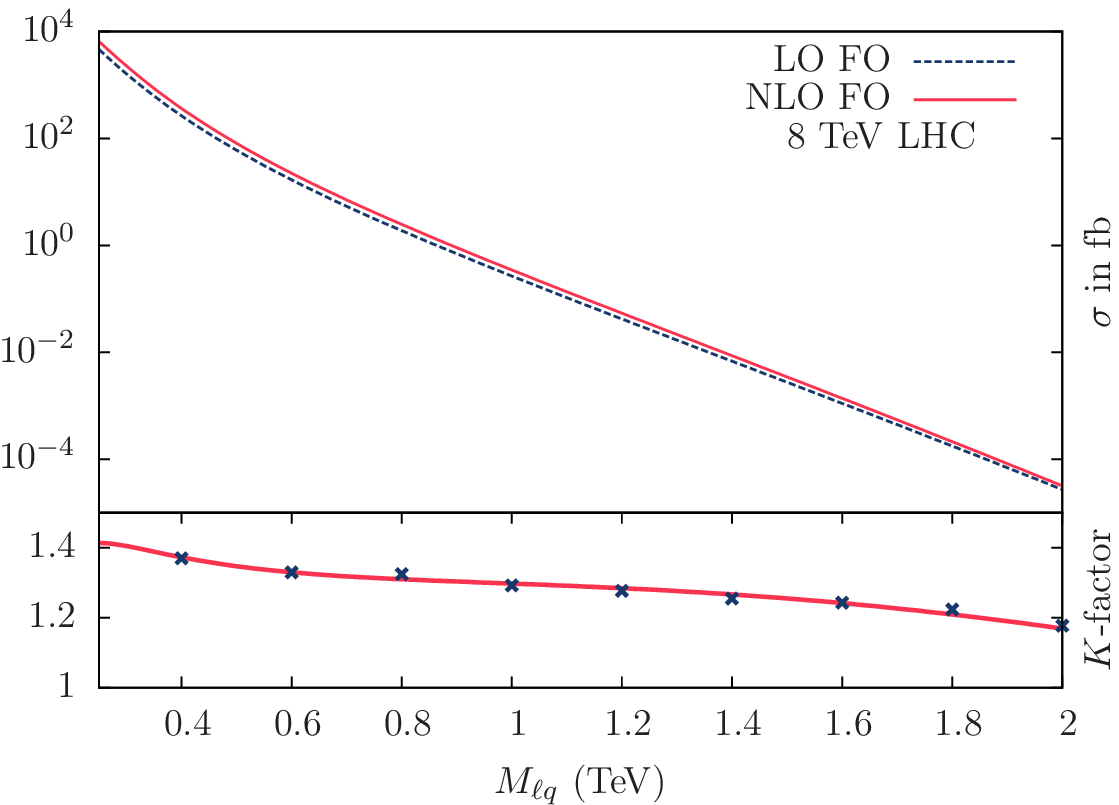}}
&&
\subfloat[]{\includegraphics[width=\linewidth]{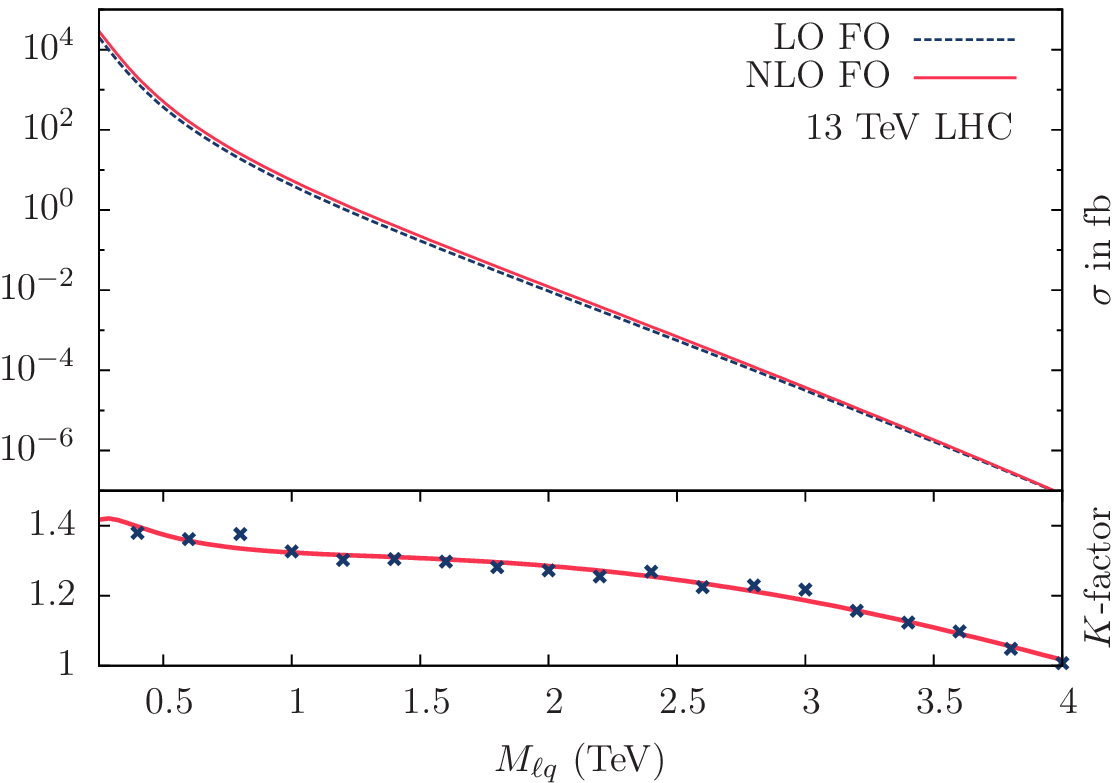}}\\
\subfloat[]{\includegraphics[width=\linewidth]{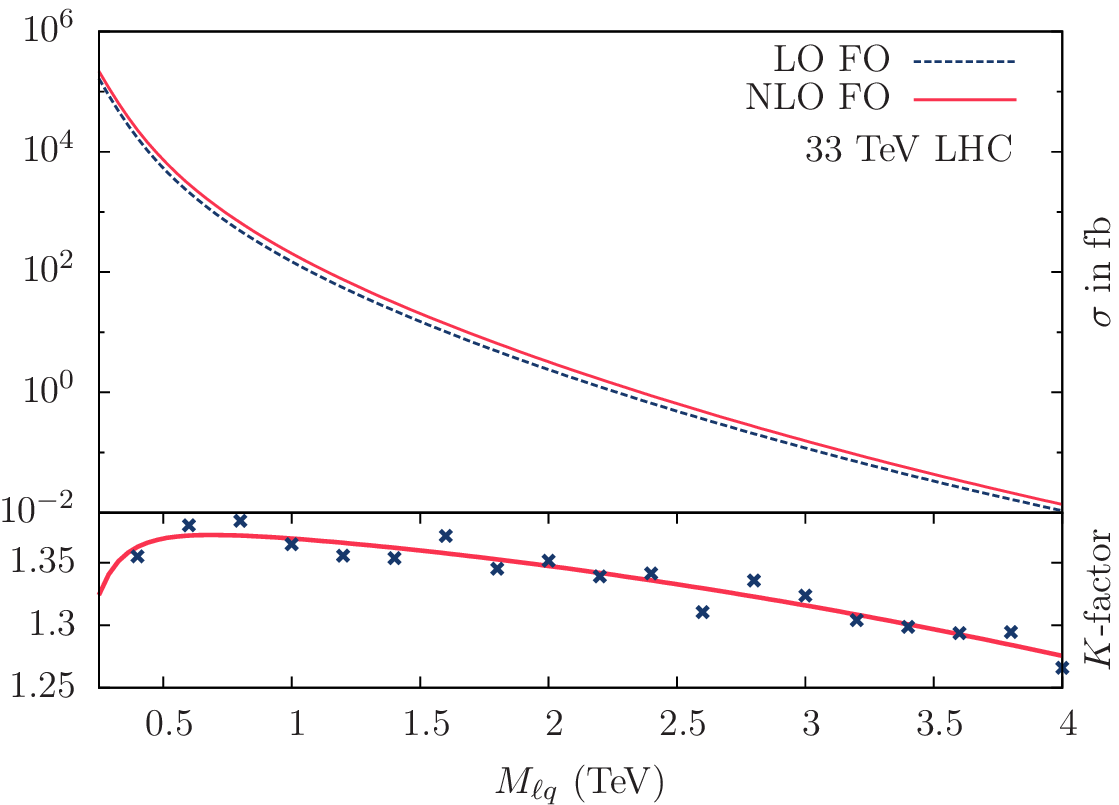}} 
&&
\subfloat[]{\includegraphics[width=\linewidth]{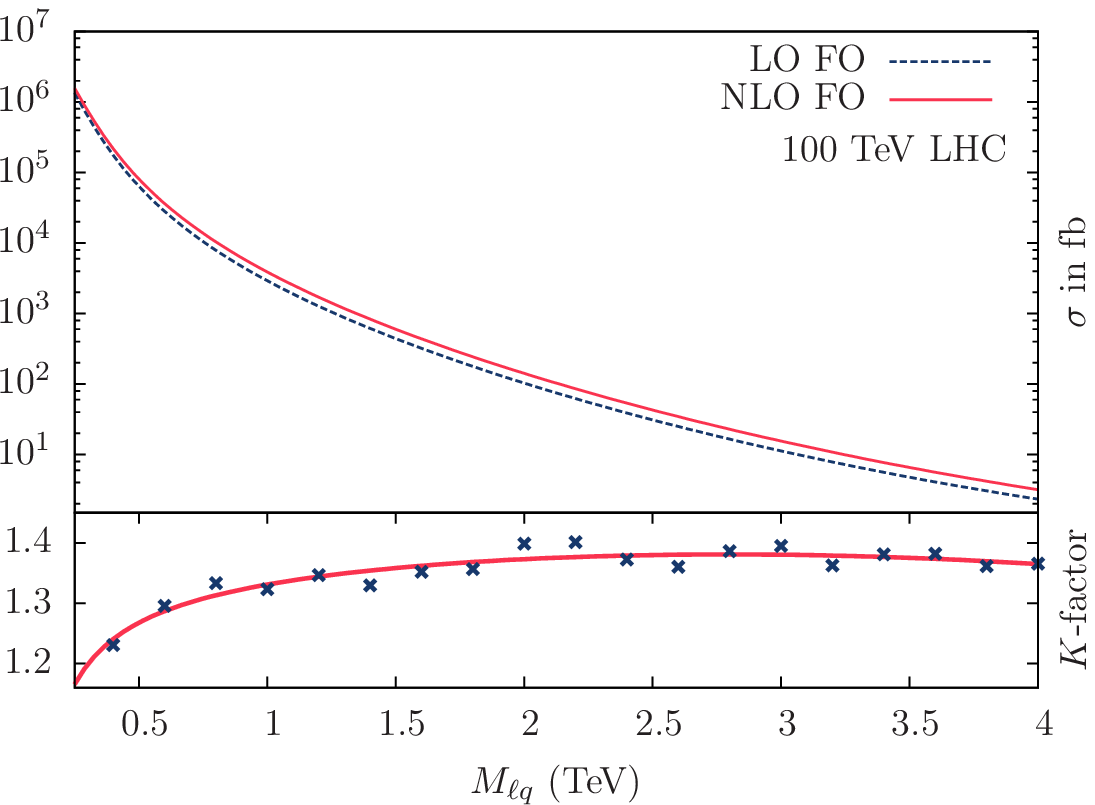}}
\end{tabular}\ec
\caption{The LO and NLO cross sections computed with MSTW(n)lo2008cl68 PDFs \cite{Martin:2009iq} and $\m_F=\m_R=M_{\lq}$  for four different
center-of-mass energies at the LHC environment -- (a) 8 TeV, (b) 13 TeV, (c) 33 TeV and (d) 100 TeV. The corresponding $K$-factors are shown in the lower panels. The lines represent the fitting functions defined in Eq.~\eqref{eq:fit} with the coefficients given in Table~\ref{tab:xsec} whereas the points are obtained from direct computations. }
\label{fig:fit}
\end{figure*}

\begin{table}
\bc
\begin{tabular}{|lllllll|}
\hline $\sqrt S$ 	& QCD	order	&	$C_{-2}$	&$C_{-1}$	&$C_{0}$	&$C_1$		&$C_2$\\\hline\hline
8~TeV							 	& LO		& -0.209		&~2.655			&~2.597		& -5.534		&-0.831	\\
								&NLO		&-0.227		&~2.814			&	~2.524	&-5.226		&-0.946    \\\hline
13~TeV							 	& LO		&-0.283		&~3.184			&~2.708		&-3.931		&-0.256	\\
								&NLO		&-0.300		&~3.318			&	~2.762	&-3.780		&-0.299    \\\hline					
33~TeV							 	& LO		& -0.217		&~2.581			&~6.059		&-3.617		&~0.203	\\
								&NLO		&-0.221 		&~2.590 		&	~6.377 	& -3.621  	&~0.199     \\\hline		
100~TeV							 	& LO		&-0.196		&~2.343			&~8.613		&-3.014		&~0.232	\\
								&NLO		& -0.196		&~2.307 		&	~8.909 	& -2.981  	&~0.225      \\\hline											
\end{tabular}
\caption{Coefficients of the cross section fitting functions defined in Eq.~\eqref{eq:fit} for different center-of-mass energies $(\sqrt S)$ at the LHC computed with MSTW(n)lo2008cl68 PDF sets \cite{Martin:2009iq}.}\label{tab:xsec}
\ec
\end{table}

\begin{figure}[!t]
\centering
\includegraphics[width=0.8\linewidth]{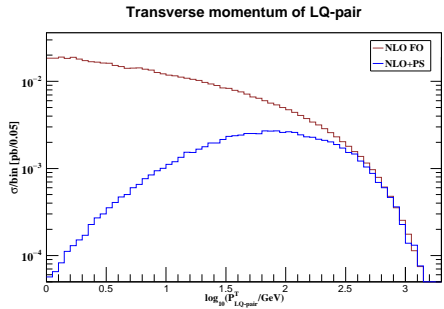}
\caption{Transverse momentum distribution of the LQ-pair at fixed order NLO and NLO+PS at the 14 TeV LHC.}
\label{logpt}
\end{figure}

  \begin{figure*}
\bc
\begin{tabular}{m{0.4\linewidth}m{0.4\linewidth}}
\subfloat[]{\includegraphics[width=\linewidth]{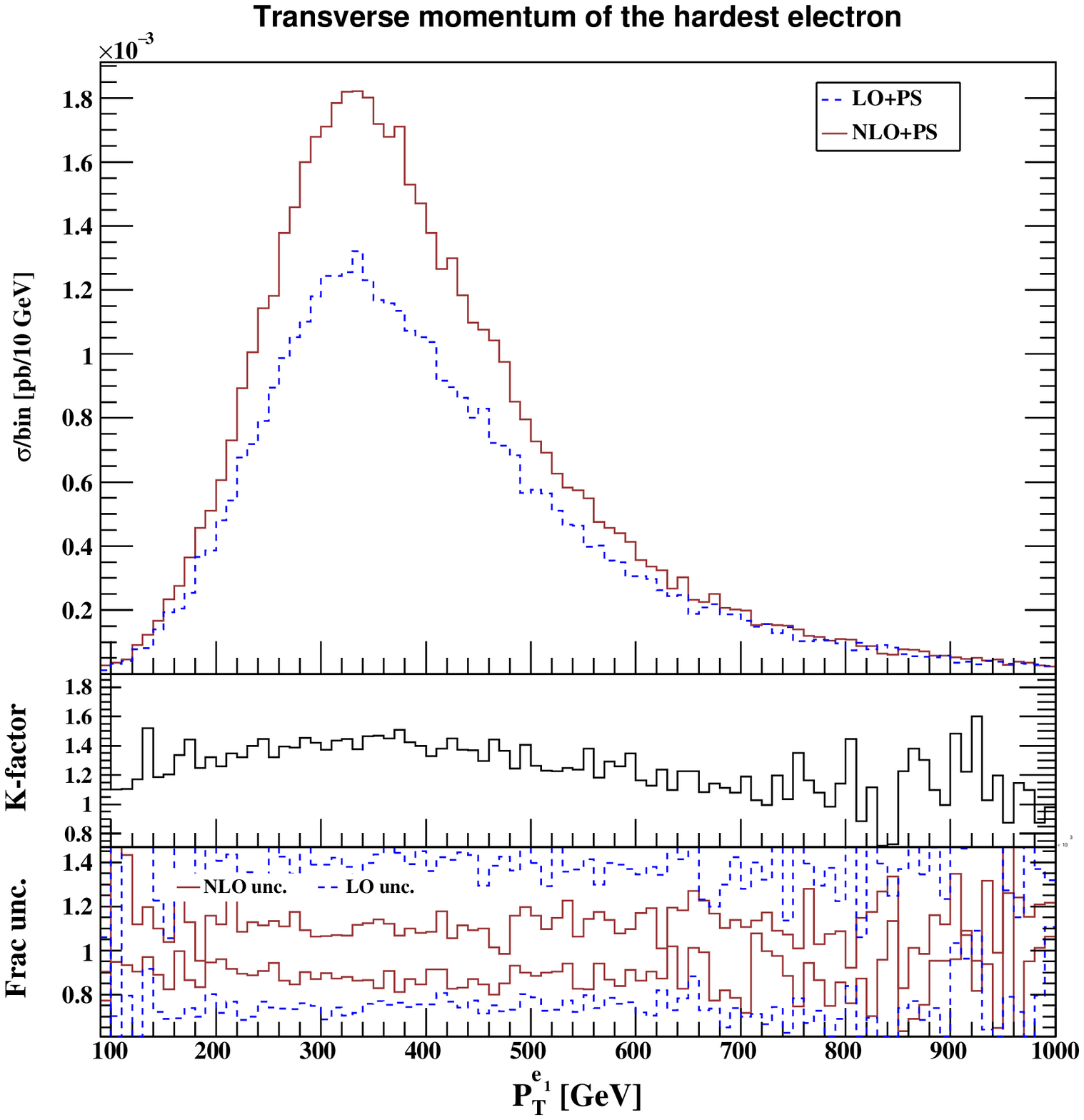}\label{fig:eejj_pt_e1}} 
&
\subfloat[]{\includegraphics[width=\linewidth]{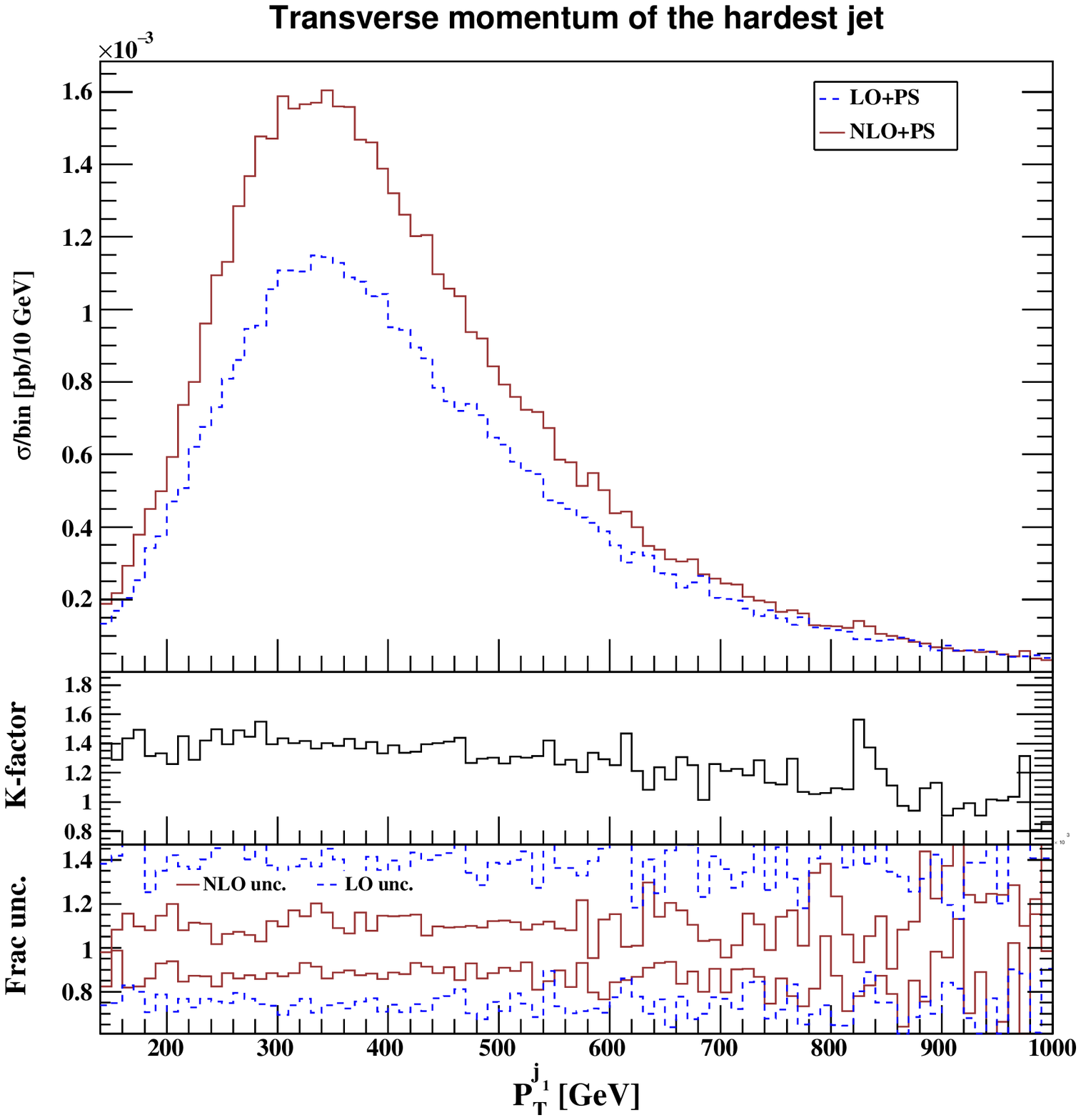}\label{fig:eejj_pt_e2}}\\
\subfloat[]{\includegraphics[width=\linewidth]{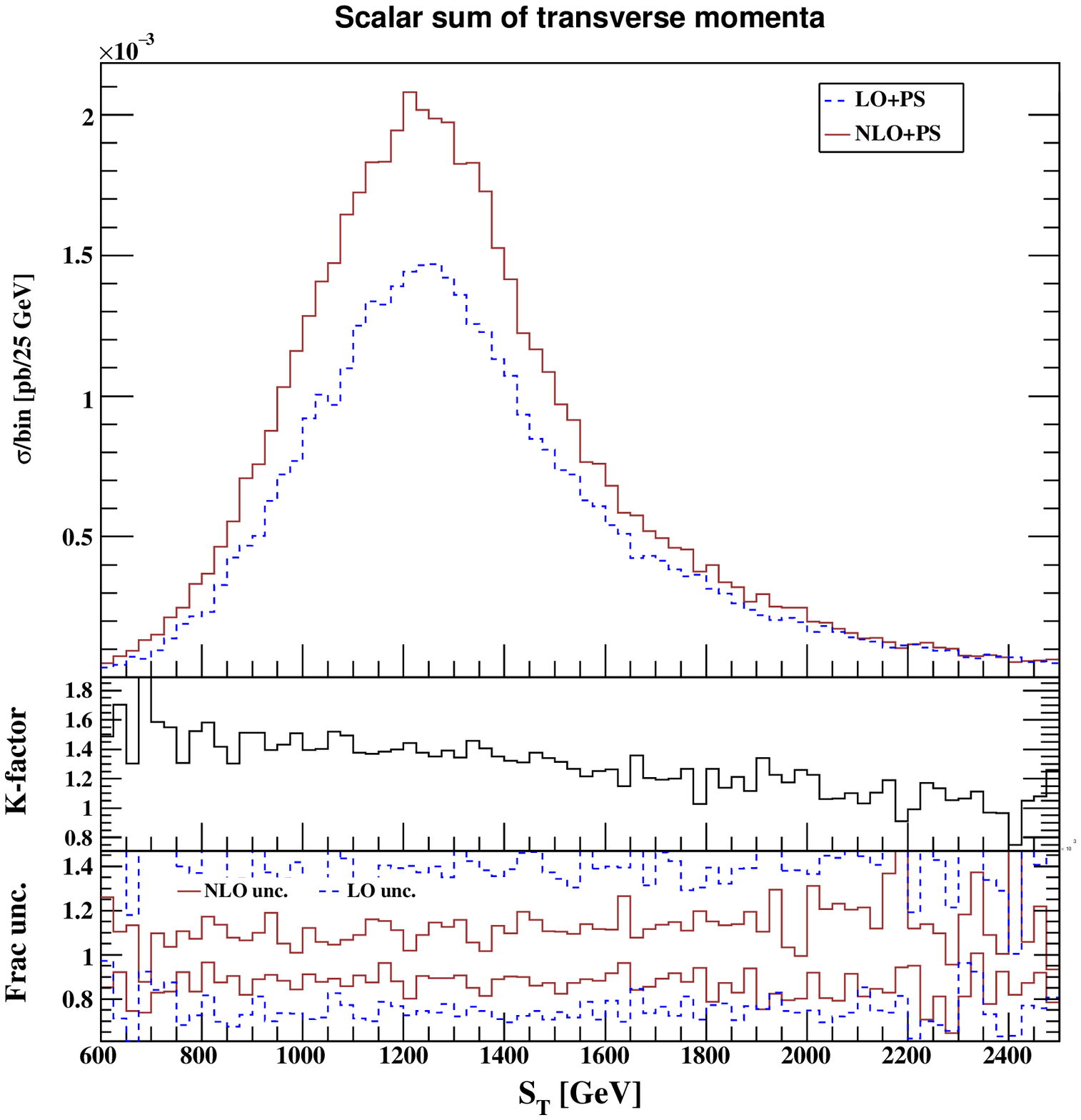}\label{fig:eejj_st_cms}} 
&
\subfloat[]{\includegraphics[width=\linewidth]{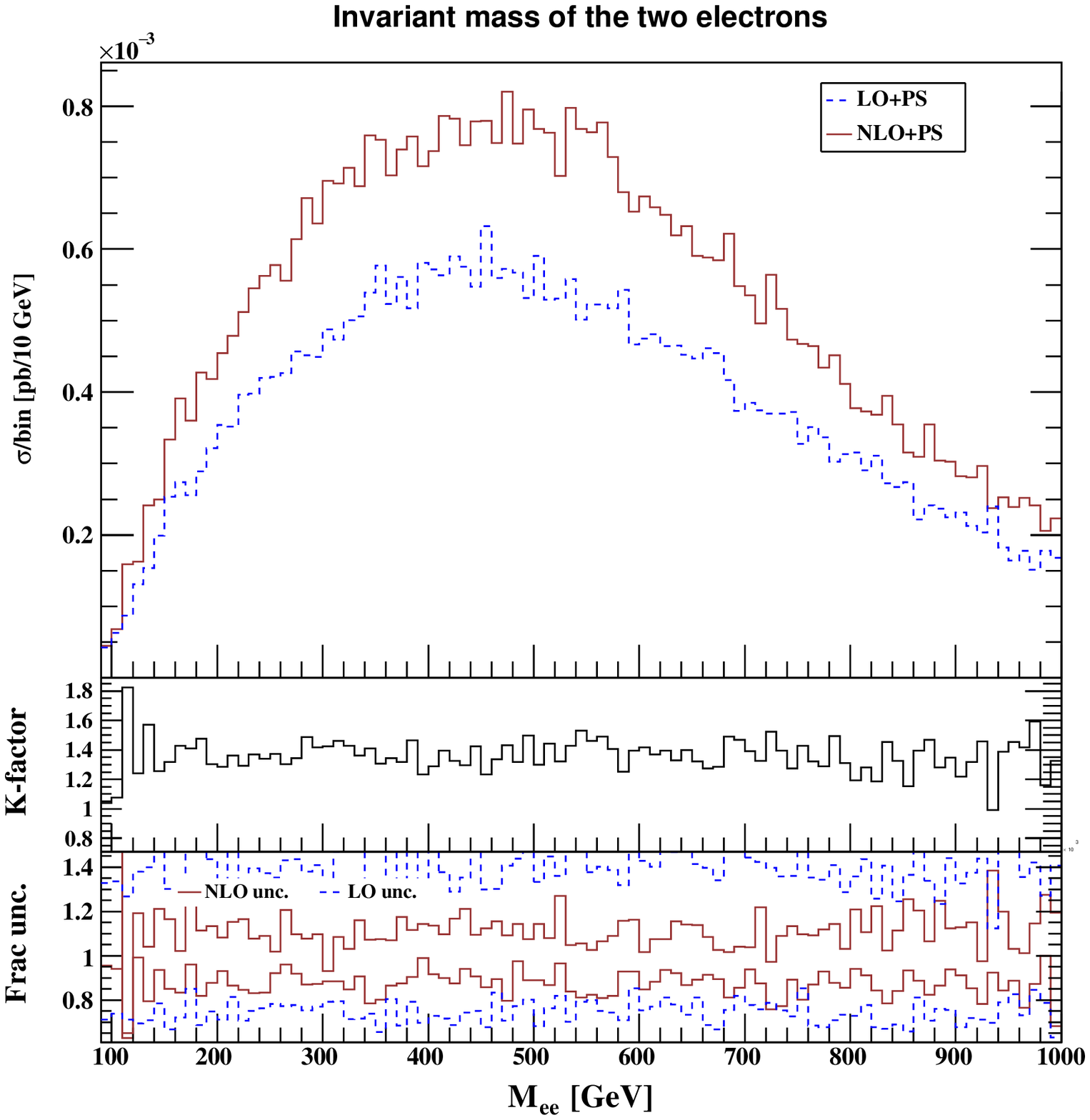}\label{fig:eejj_inv_e1e2}} \\
\subfloat[]{\includegraphics[width=\linewidth]{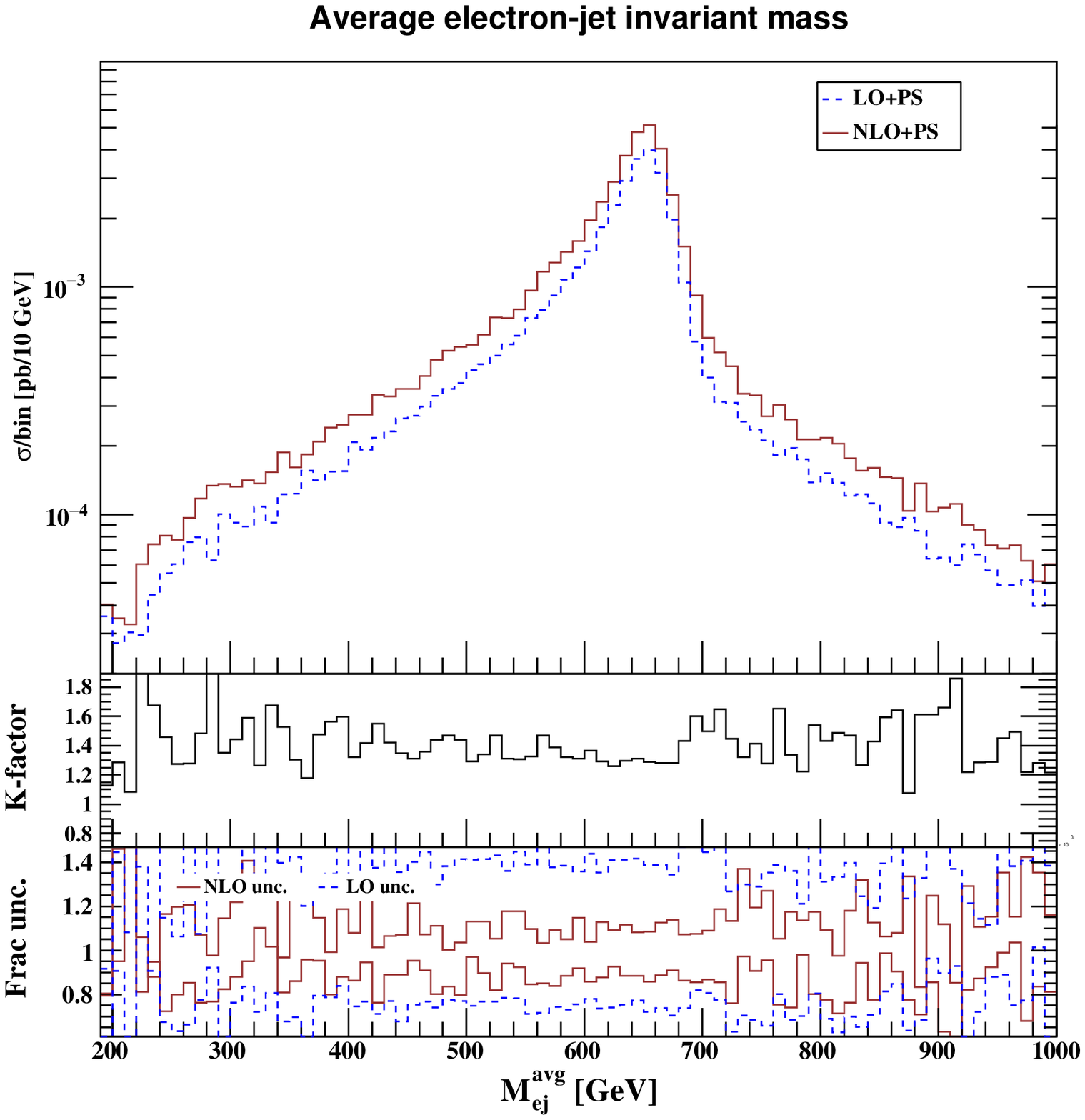}\label{fig:eejj_inv_ejavg}} 
&
\subfloat[]{\includegraphics[width=\linewidth]{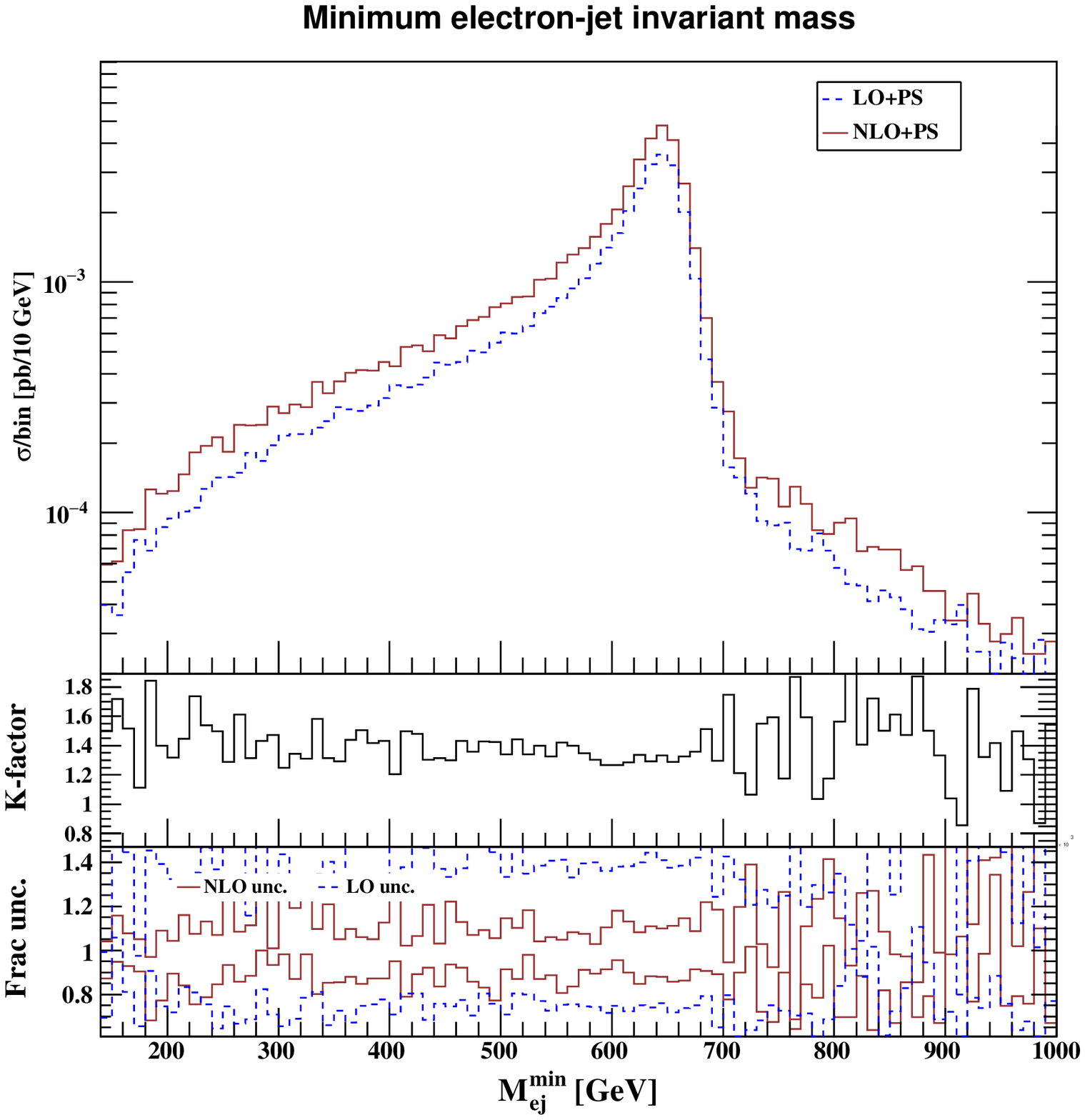}\label{fig:eejj_inv_ejmin}}\\
\end{tabular}\ec
\caption{Kinematic distributions for the $eejj$ channel. For these plots we set $M_{\lq} = 650$ GeV and $\bt =1$.}
\label{fig:eejjDists}
\end{figure*}

\begin{figure*}
\bc
\begin{tabular}{m{0.4\linewidth}m{0.4\linewidth}}
\subfloat[]{\includegraphics[width=\linewidth]{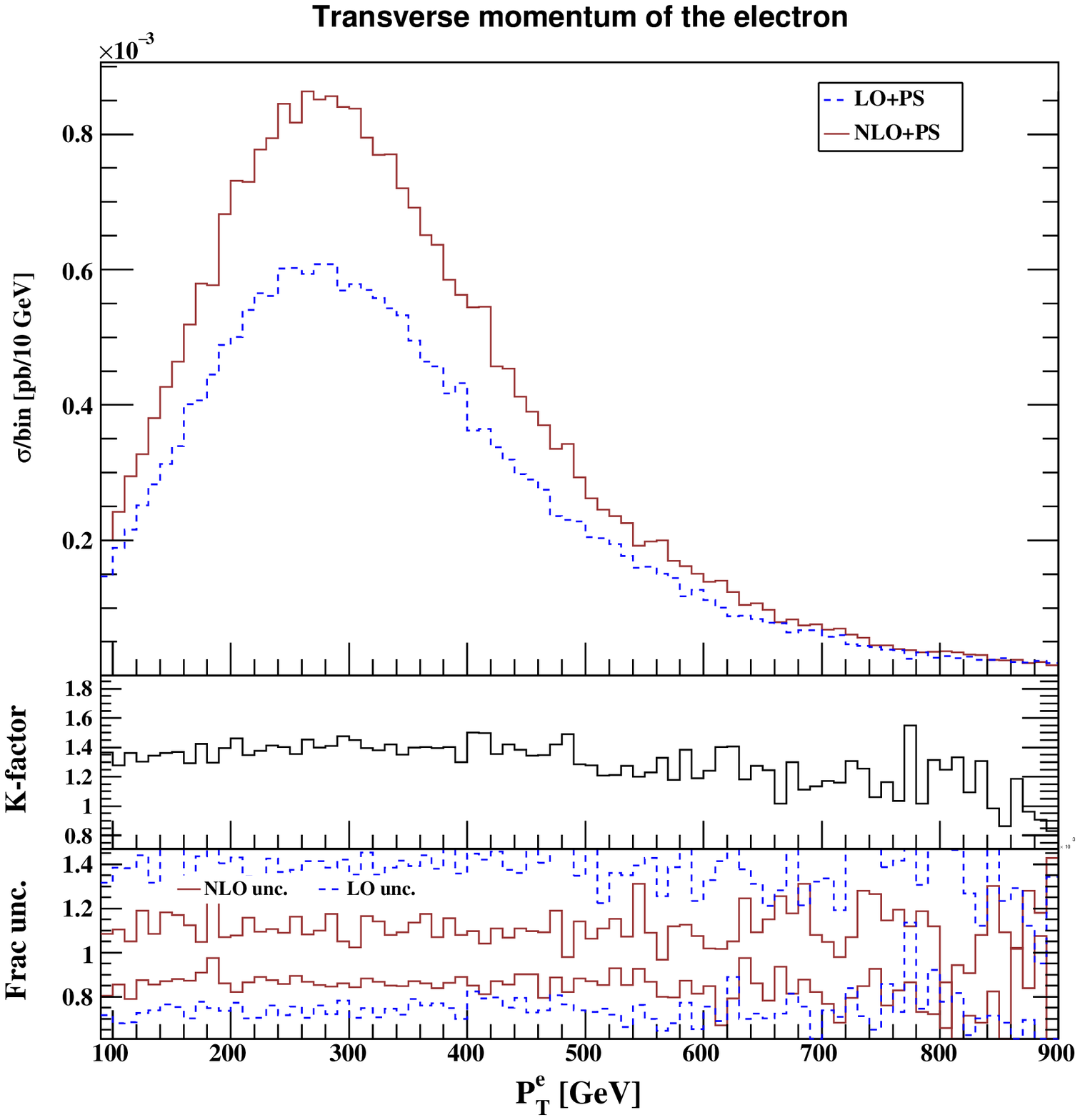}\label{fig:evjj_pt_e1}} 
&
\subfloat[]{\includegraphics[width=\linewidth]{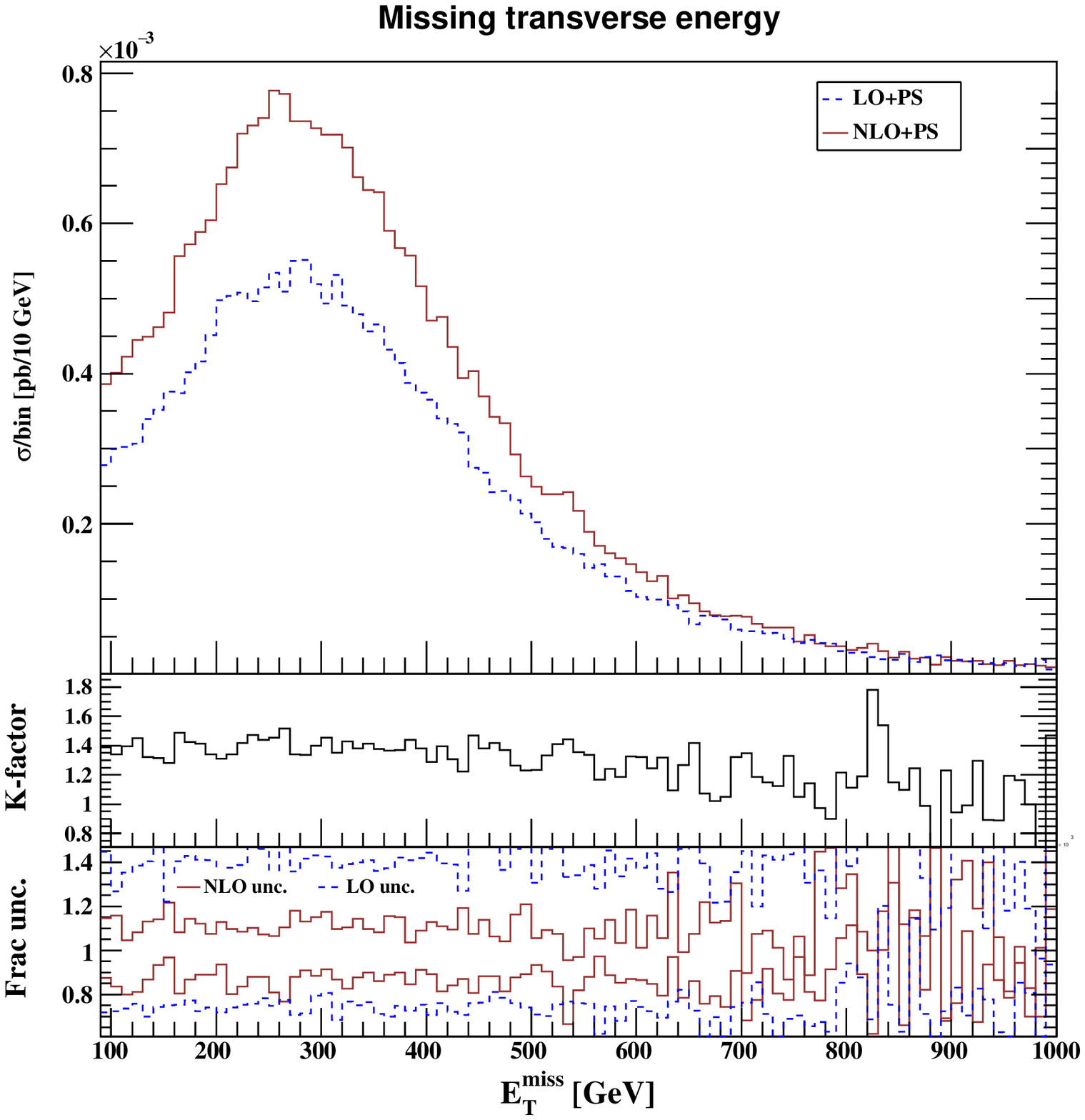}\label{fig:evjj_MET}}\\
\subfloat[]{\includegraphics[width=\linewidth]{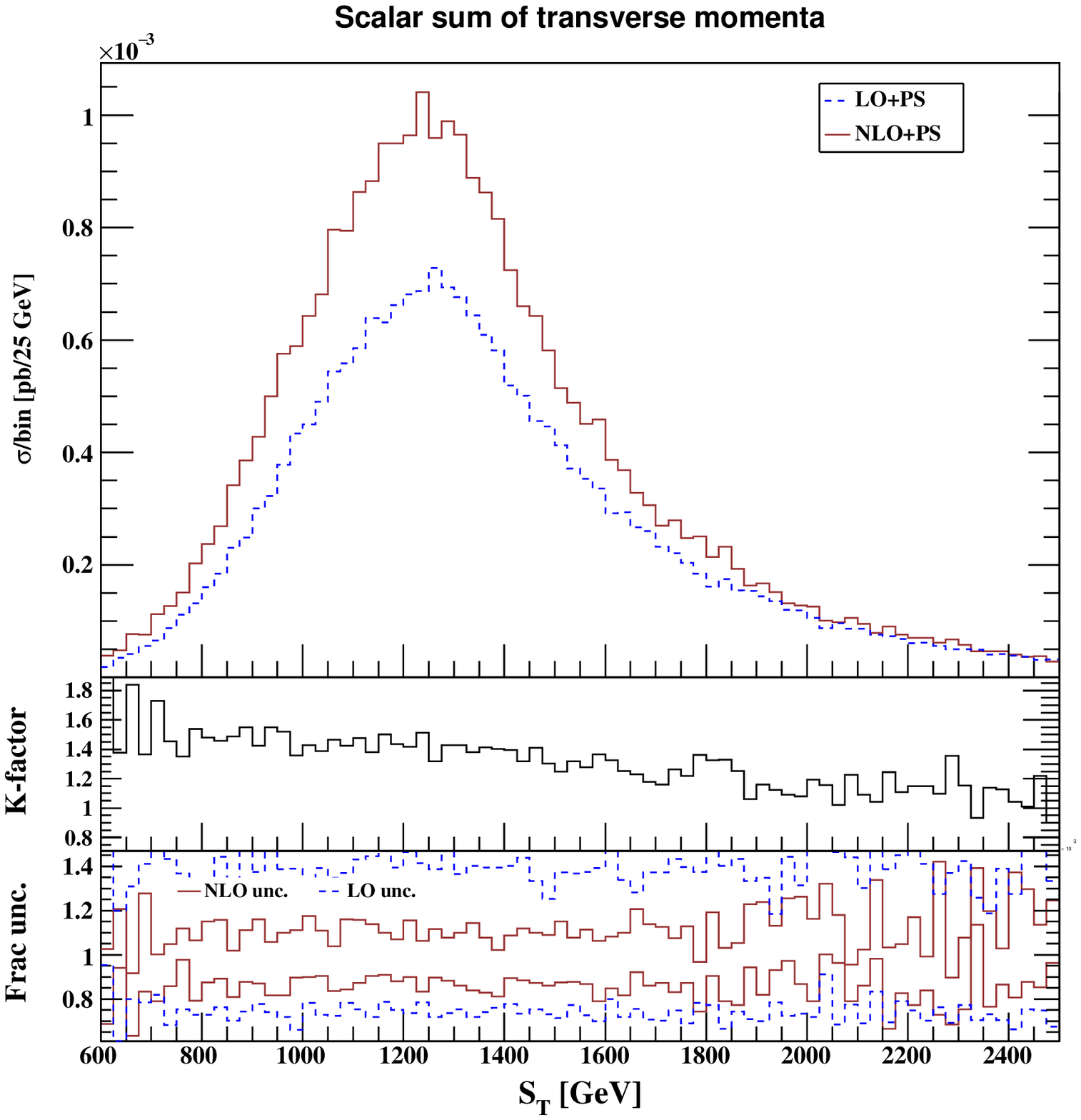}\label{fig:evjj_st_cms}} 
&
\subfloat[]{\includegraphics[width=\linewidth]{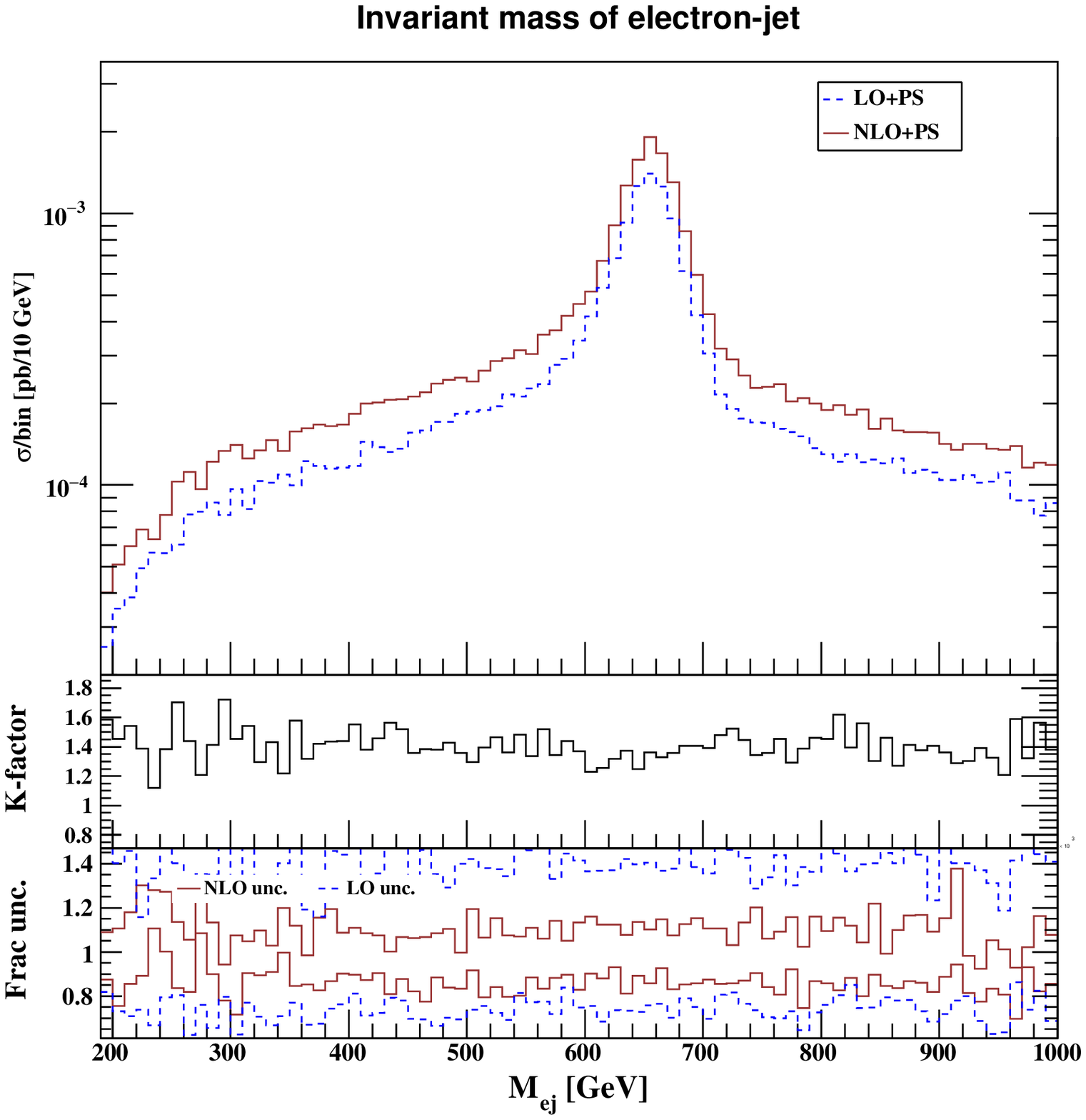}\label{fig:evjj_inv_ej}}  \\
\subfloat[]{\includegraphics[width=\linewidth]{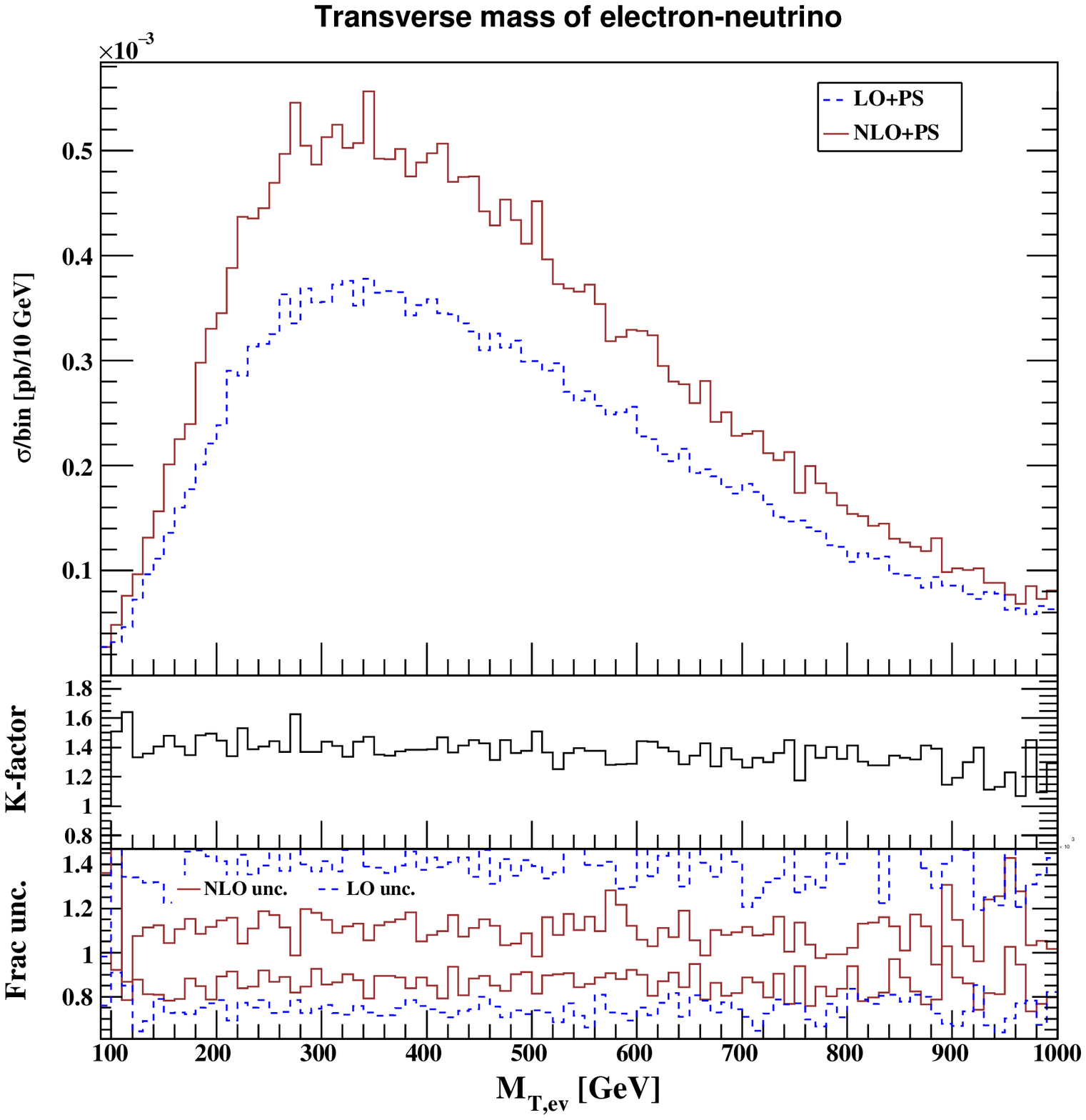}\label{fig:evjj_tmass_ev}}
&
\subfloat[]{\includegraphics[width=\linewidth]{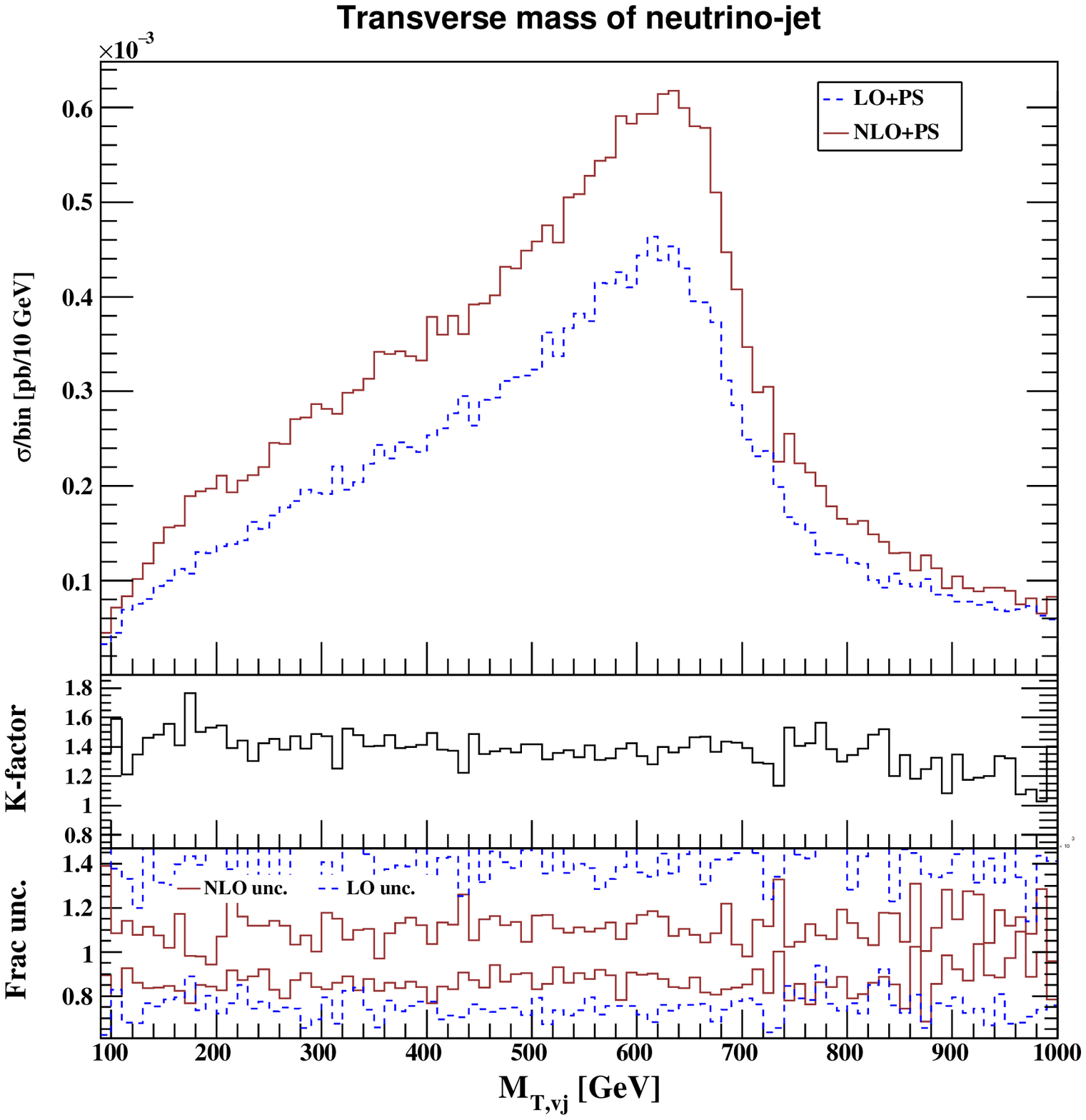}\label{fig:evjj_tmass_vj}}
\end{tabular}\ec
\caption{Kinematic distributions for the $e\n jj$ channel. For these plots we set $M_{\lq} = 650$ GeV and $\bt =0.5$.}
\label{fig:evjjDists}
\end{figure*}

\begin{figure*}
\bc
\begin{tabular}{m{0.45\linewidth}m{0.45\linewidth}}
\subfloat[]{\includegraphics[width=\linewidth]{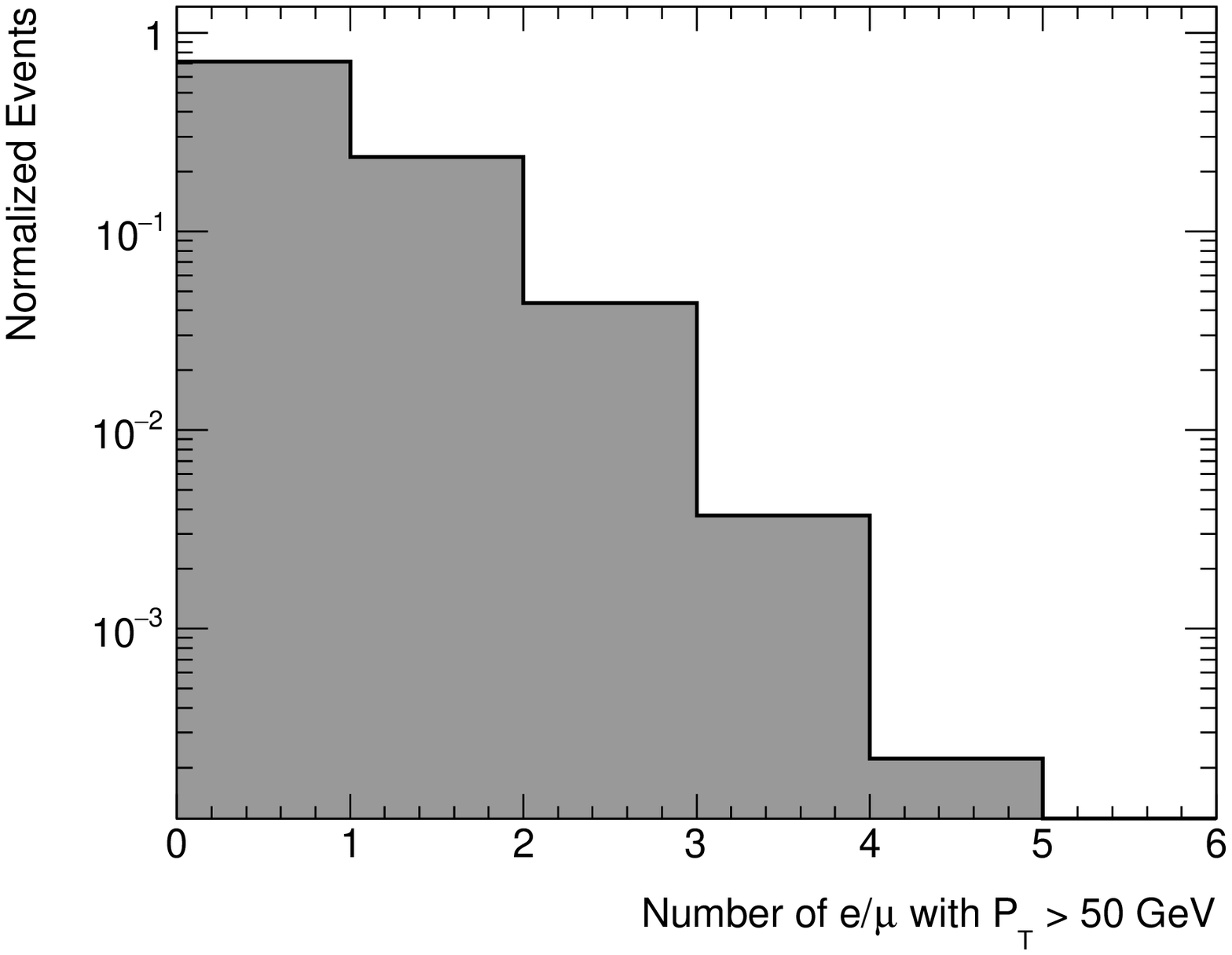}\label{fig:3g_lep_no}} 
&
\subfloat[]{\includegraphics[width=\linewidth]{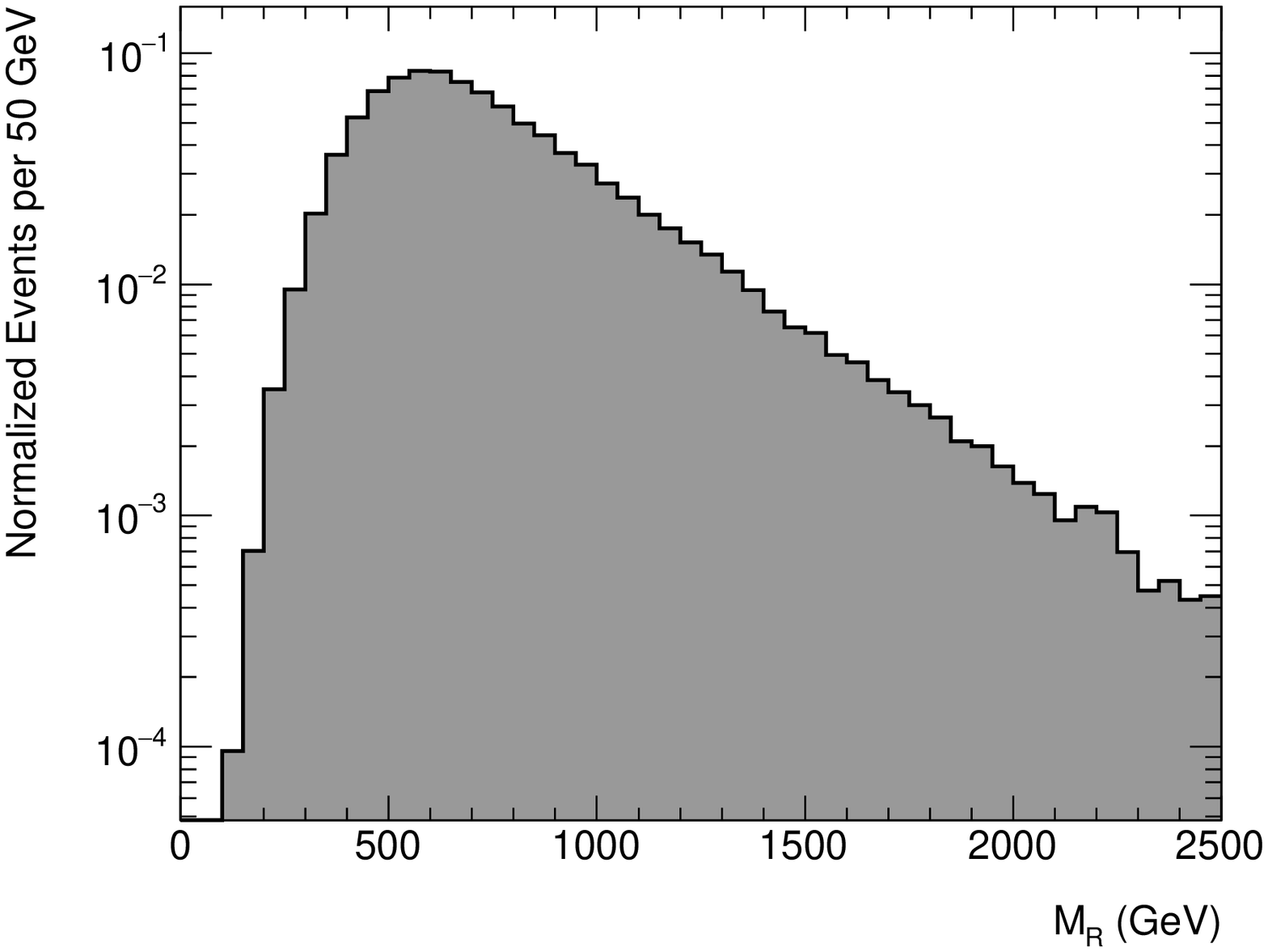}\label{fig:3g_razor}}
\end{tabular}\ec
\caption{Distributions for third generation charge 1/3 LQs. For these plots, we set $M_{\ell_q} = 650$ GeV and $\beta = 0.5$. See Eq. \eqref{eq:MR} for the definition of $M_{\rm R}$.}
\label{fig:3g}
\end{figure*}

In Table~\ref{tab:compare}, we compare the FO NLO cross sections and the $K$-factors 
for 14 TeV LHC computed with CTEQ6(M/L1) PDFs \cite{Pumplin:2002vw} and $n_f=5$ massless quark flavors (i.e., the same 
scheme as in Ref.~\cite{Kramer:2004df}) with those obtained with our scheme. The numbers obtained with CTEQ PDFs can be 
compared directly with the ones shown in Table I of Ref.~\cite{Kramer:2004df}.

In Fig.~\ref{fig:fit}, we show the LQ pair production ($pp\to\lq\bar\lq$) cross sections computed with  $\zeta=1$ [Eq.~\eqref{eq:scales}] for four different center-of-mass energies ($\sqrt S$) at the LHC environment (8, 13, 33 and 100 TeV). The corresponding $K$-factors are also shown in the lower panels.  The cross sections (LO and NLO) can be fitted with the following fitting function to a very good accuracy,
\ba
\sg \mbox{ (in fb)}\bigg|_{\m_F=\m_R=M_{\lq}}  &\approx& \exp\lt\{\sum_{n=-2}^2 C_{n} \lt(\frac{M_{\lq}}{\rm TeV}\rt)^n\rt\}\,,\nn\\\label{eq:fit}
\ea
where the coefficients are given in Table~\ref{tab:xsec}. These fits are valid for $M_{\lq}$ between 0.25  and 4 TeV for all the center-of-mass energies except $\sqrt{S}=8$ TeV for which the range of validity is $0.25$ TeV $\leq M_{\lq} \leq 2$ TeV.

To show the effect of the PS over the NLO FO  results, we plot the transverse momentum distribution, 
$\log_{10}(p_{\rm T}^{\rm{LQ-pair}})$, of the LQ-pair in Fig.~\ref{logpt} for the 14 TeV LHC.  
The fixed order (NLO FO) distribution is shown by the dashed blue line, and the NLO matched with the parton shower (NLO+PS) result is shown with a solid brown line. The NLO FO  result is 
divergent as $p_{\rm T}\to 0$ and therefore it is necessary to match it with PS in order to get a reliable estimation of the cross section in this region. The converging behavior of 
the NLO+PS distribution in the low $p_{\rm T}$ region indicates the correct resummation of the Sudakov logarithms in the collinear region, 
thereby leading to a notable Sudakov supression visible prominently in the log-log plot of Fig.~\ref{logpt}.
As expected, both of these results agree in the high $p_{\rm T}$ region. 

\renewcommand\theenumi{\emph{(\roman{enumi})}}
\renewcommand\labelenumi{\theenumi}

Since the analyses for the first two generations of LQs are very similar, in the next two figures, we present results of kinematic distributions of a selection of observables at (N)LO+PS accuracy only for the first generation. Note that, from here onward, we set the typical value of LHC center-of-mass energy to $\sqrt{S}=13$ TeV unless specified otherwise and pass the {\sc Pythia} events through the detector simulator {\sc Delphes 3.3.1}~\cite{deFavereau:2013fsa} with the default CMS card to generate the distributions. 
To obtain the distributions, we apply the  basic preselection kinematic cuts from the CMS analysis\footnote{These cuts are designed for the 8 TeV LHC. However, since they are only for the  preselection, we use them to get the 13 TeV distributions for easy comparison. }~\cite{CMS:2014qpa}  to the $eejj$ final states:
\begin{enumerate}
 \item exactly two electrons ($e^{\pm}$)
with transverse momentum $p_{\rm T}^e > 45$ GeV and pseudorapidty $|\eta_e| < 2.5$ excluding  $1.442 < |\eta_e| < 1.56$,
 \item two leading $p_{\rm T}$ jets with $p_{\rm T}^{j_1} > 125$ GeV, $p_{\rm T}^{j_2} > 45$ GeV and $|\eta_j| < 2.4$, 
 \item separation between an electron and a jet in the $\eta-\phi$ plane, $\Delta R_{ej} > 0.3$ 
 \item the invariant mass of the electron pair, $M_{ee} > 50$ GeV, 
 \item the scalar sum of the $p_{\rm T}$ of the two electrons and the two leading $p_{\rm T}$ jets, $S_{\rm T} > 300$ GeV, 
 \item the minimum of electron-jet invariant mass combinations, $M_{ej}^{\rm min} \geq 50$ GeV, where, to form two $ej$-pairs out of $eejj$ final states, 
combination with the smaller difference between the two $ej$ invariant masses ($M_{ej}$)  is considered. 
\end{enumerate}
Similarly, for the $e\n jj$ final states, we apply
\begin{enumerate}
 \item exactly one electron with $p_{\rm T}^e > 45$ GeV and $|\eta_e| < 2.1$ excluding  $1.442 < |\eta_e| < 1.56$,
 \item two leading $p_{\rm T}$ jets with $p_{\rm T}^{j_1} > 125$ GeV, $p_{\rm T}^{j_2} > 45$ GeV and $|\eta_j| < 2.4$, 
 \item the missing transverse energy, $E_{\rm T}^{\rm miss} > 55$ GeV, 
 \item azimuthal separation between the electron and the $E_{\rm T}^{\rm miss}$, $\Delta\phi(e,E_{\rm T}^{\rm miss}) > 0.8$, 
 \item azimuthal separation between the hardest jet and the $E_{\rm T}^{\rm miss}$, $\Delta\phi(j_1,E_{\rm T}^{\rm miss}) > 0.5$, 
 \item separation between the electron with the two leading $p_{\rm T}$ jets in the $\eta-\phi$ plane, $\Delta R_{ej} > 0.7$, 
 \item the scalar sum of the $p_{\rm T}$ of the electron, the $E_{\rm T}^{\rm miss}$ and the two leading jets, $S_{\rm T} > 300$ GeV,
 \item the electron-neutrino transverse mass, $M_{{\rm T}, e\n}=\sqrt{2p_{\rm T}^eE_{\rm T}^{\rm miss}\left[1-\cos\left(\Delta\ph\left(E_{\rm T}^{\rm miss}, e\right)\right)\right]}>50$ GeV, 
 \item the electron-jet invariant mass, $M_{ej} > 50$ GeV, in which the electron and the jet are picked out of that combination for which 
the difference between the electron-jet transverse mass and the neutrino-jet transverse mass is smaller. 
\end{enumerate}
To present the plots, we adopt a systematic graphical representation scheme: the main frame shows the kinematical distributions of an observable at 
LO+PS (dashed blue) and NLO+PS (solid brown) accuracy, the middle pad represents the ratio ($K$-factor) of the NLO+PS result over the LO+PS result 
with a solid black line and in the lower inset, and we present the fractional scale uncertainties of corresponding observables [dashed blue (solid brown) for (N)LO+PS], defined 
by the ratio of their variation around the central scale. 

In Fig.~\ref{fig:eejjDists}, we show the distributions for the $eejj$ channel. Our choice of  variables is mainly motivated by the CMS analysis 
\cite{CMS:2014qpa} (ALTAS also uses similar variables~\cite{Aad:2011ch}). In addition to the $p_{\rm T}$ of the hardest electron and jet
[Figs.~\ref{fig:eejj_pt_e1} and \ref{fig:eejj_pt_e2}], we show the distributions of $S_{\rm T}$, $M_{ee}$, $M_{ej}^{\rm avg}$ (the average of electron-jet invariant mass combinations) and 
$M_{ej}^{\rm min}$ in Figs.~\ref{fig:eejj_st_cms}, \ref{fig:eejj_inv_e1e2}, \ref{fig:eejj_inv_ejavg} and \ref{fig:eejj_inv_ejmin} respectively.
For the $e\n jj$ channel, we show the distributions in Fig.~\ref{fig:evjjDists}:
({\em i}) $p_{\rm T}^e$ [Fig.~\ref{fig:evjj_pt_e1}],
({\em ii}) $E_{\rm T}^{\rm miss}$ [Fig.~\ref{fig:evjj_MET}],
({\em iii}) $S_{\rm T}$ [Fig.~\ref{fig:evjj_st_cms}],
({\em iv}) $M_{ej}$ [Fig.~\ref{fig:evjj_inv_ej}],
({\em v}) $M_{{\rm T}, e\n}$
[Fig.~\ref{fig:evjj_tmass_ev}], and
({\em vi}) the neutrino-jet transverse mass $M_{{\rm T}, \n j}=\sqrt{ (E_{\rm T}^{\rm miss} + E_{\rm T}^j)^2 - (\vec{p}_{\rm T}^{\;\rm miss} + \vec{p}_{\rm T}^{\;j})^2}$ [Fig.~\ref{fig:evjj_tmass_vj}]
where the combination with
the smaller difference between the electron-jet transverse mass and the neutrino-jet
transverse mass was considered.

For all the kinematical observables presented in Figs.~\ref{fig:eejjDists} and \ref{fig:evjjDists}, we observe that the NLO+PS distributions 
are substantially larger than the corresponding LO+PS results and the scale uncertainties are notably reduced at NLO+PS. 
Though not shown, we find 
no significant differences between the PDF uncertainties computed at the NLO+PS and LO+PS levels using Hessian method as suggested by the MSTW Collaboration~\cite{Martin:2009iq}.

In Figs.~\ref{fig:eejj_pt_e1} and \ref{fig:eejj_pt_e2} or~\ref{fig:evjj_pt_e1} and \ref{fig:evjj_MET} we find that in the high $p_{\rm T}/E_{\rm T}^{\rm miss}$ regions, there is no notable difference between NLO+PS and LO+PS; i.e., the $K$-factor is close to 1 in those regions. This is because, while the $E_{\rm T}^{\rm miss}$ or the $p_{\rm T}$ of an electron or the leading jet increases, the extra emission at NLO tends to have smaller $p_{\rm T}$ as  
the center of mass energy remains unaltered. A similar argument holds for Fig.~\ref{fig:eejj_st_cms} or.~\ref{fig:evjj_st_cms}. In the other distributions, the $K$-factor is nearly constant in the regions shown; however with the increase in mass scale it comes down to 1 as it should. 

For the third generation, as already mentioned,  we have implemented both Model A and Model B in our code. Here, however, we only display the plots for Model A type LQs that decay either to $t$-quarks and $\ta$'s or to $b$-quarks and $\n_\ta$'s. In Fig.~\ref{fig:3g_lep_no}, we show the number of hard (with $p_{\rm T}>50$ GeV) electrons or muons in the events generated with $\bt=0.5$. Notice, in this case there can be more than two charged leptons in the final states from decays like
\ba
 \lq \to t \ta\ubr{-2.0} \to \underbrace{(b\ \hat\ell\ \n_{\hat\ell})}_{t{\rm - decay}}\ \overbrace{(\hat\ell\ \n_\ta\ \n_{\hat\ell})}^{\ta{\rm -decay}}
\ea
where the two light leptons ($e$ or $\m$) coming from the same LQ will have opposite charges. For a LQ with electric charge $\pm4/3$, they would carry the same charge but such a LQ would not couple to a $\n$ and a $b$-quark i.e., $\bt$ cannot be $0.5$ (as we have set here), in the absence of any unknown decay mode.
This multilepton signature is unique for the third generation, as pair productions of the first two generations of LQs can produce only two hard leptons in the final state.

For the events where the LQs have decayed into $b$-quarks and $\n_\ta$'s we have plotted $M_R$ (see Fig.~\ref{fig:3g_razor}), defined as
\ba
M_R = \sqrt{\left(|\vec p^1|^2 + |\vec p^2|\right)^2 - (p^1_z+p^2_z)^2},\label{eq:MR}
\ea
where $p^i$ refers to the $i$th $p_{\rm T}$ ordered jet and $p^i_z$ its longitudinal component. It is an approximation of the razor mass used in Ref.~\cite{Chatrchyan:2012st}, as we have used the first two hard jets instead of ``pseudojets'' (see Ref.~\cite{Chatrchyan:2012st} for details). It is suitable for identifying a heavy resonance decaying into a jet and a weakly interacting neutral particle from the SM background. For our demonstration plot, we have selected events with no light charged lepton and $E_{\rm T}^{\rm miss}>140$ GeV.
\section{Conclusions}

 LQs were hypothesized a few decades back but mainly because of the failures of earlier LQ-searches at HERA and Tevatron, nowadays they receive much less attention than other BSM candidates. However,  they are still relevant today (see e.g., Refs.~\cite{Hiller:2014yaa,Freytsis:2015qca,Bauer:2015knc}) and the LHC is actively looking for their signatures.

In this paper, we have taken another look at the NLO QCD corrections to the scalar LQ pair production process at the LHC and included the parton shower effects to obtain new NLO+PS accurate results. Our computation provides a reliable and completely automated code that goes beyond computing the NLO $K$-factor by simulating signal events at the QCD NLO(+PS) level and improves the precision of various kinematic distributions. These precise distributions can play a crucial role in advanced techniques like multivariate analysis. The excellent agreement between our cross section computations with the available results demonstrates the reliability of our code. It is prepared using the {\sc MadGraph}5\_{\sc aMC@NLO} framework and therefore, is flexible to the choice of energy, cuts, scales, PDFs, etc. We stress that our computations cover all three generations. The signatures of the pair productions of the first two generations of LQs are similar and mostly model independent, whereas for the third generation, they are different in nature. Our code can be used for LQs decaying to either a top quark or a bottom quark in association with a third generation lepton. The complete stand-alone code is available publicly at the website -- {\url{http://amcatnlo.cern.ch}}. In addition, we have also provided $K$-factors for a wide range of $M_{\lq}$ by means of fitting functions for four different LHC center-of-mass energies, $\sqrt{S}=$8, 13, 33, 100 TeV.

Finally, we note that sometimes searches for other BSM particles with similar final states are reinterpreted for LQ parameters. For example, in Ref.~\cite{Chatrchyan:2012st}, a search for particles decaying to a $b$-quark + $E_{\rm T}^{\rm miss}$ final state is interpreted in terms of $b$-squarks decaying to $b$-quarks and  $\chi_0$'s and third generation LQs decaying to $b$-quarks and $\n_{\ta}$'s. However, this could be misleading if the $\chi_0$ has large mass. With our code, such problems can be avoided by simulating the LQ signal precisely and independently.


\section*{Acknowledgements}

T.M. is supported by funding from the DAE, for the RECAPP, HRI and the Carl Trygger Foundation under Contract No. CTS-14:206 and the Swedish Research Council under Contract No. 621-2011-5107. S.M. thanks Pankaj Jain for his encouragements and kind hospitality at IIT Kanpur where a part of this work was carried out. S.S. would like to thank R. Frederix for helpful communications.

\end{document}